\documentclass[pdflatex,sn-mathphys-num]{sn-jnl}

\usepackage{graphicx}%
\usepackage{multirow}%
\usepackage{amsmath,amssymb,amsfonts}%
\usepackage{amsthm}%
\usepackage{mathrsfs}%
\usepackage[title]{appendix}%
\usepackage{xcolor}%
\usepackage{textcomp}%
\usepackage{manyfoot}%
\usepackage{booktabs}%
\usepackage{algorithm}%
\usepackage{algorithmicx}%
\usepackage{algpseudocode}%
\usepackage{listings}%

\theoremstyle{thmstyleone}%
\theoremstyle{thmstyletwo}%

\theoremstyle{thmstylethree}%

\begin{document}

\title[Altermagnetic band splitting in 10 nm epitaxial CrSb thin films]{Altermagnetic band splitting in 10 nm epitaxial CrSb thin films}


\author[1]{\fnm{Sandra } \sur{Santhosh}}

\author[2,3]{\fnm{Paul} \sur{Corbae}}

\author[2,3]{\fnm{Wilson J.} \sur{Y\'{a}nez-Parre\~{n}o}}

\author[4]{\fnm{Supriya} \sur{Ghosh}}

\author[5]{\fnm{Christopher J.} \sur{Jensen}}

\author[6]{\fnm{Alexei V.} \sur{Fedorov}}

\author[7]{\fnm{Makoto} \sur{Hashimoto}}

\author[7]{\fnm{Donghui} \sur{Lu}}

\author[5]{\fnm{Julie A. } \sur{Borchers}}

\author[5]{\fnm{Alexander J.} \sur{Grutter}}

\author[8]{\fnm{Timothy R.} \sur{Charlton}}

\author[1,9]{\fnm{Saurav} \sur{Islam}}

\author[1,9]{\fnm{Anthony} \sur{Richardella}}

\author[4]{\fnm{K. Andre} \sur{ Mkhoyan}}

\author[2,3,10]{\fnm{Christopher J.} \sur{ Palmstr{\o}m}}

\author*[1]{\fnm{Yongxi} \sur{Ou}}\email{yongxiou@hotmail.com}

\author*[1,9,11]{\fnm{Nitin} \sur{Samarth}}\email{nxs16@psu.edu}

\affil[1]{\orgdiv{Dept. of Physics}, \orgname{Pennsylvania State University}, \city{University Park}, \postcode{16802}, \state{PA}, \country{USA}}

\affil[2]{\orgdiv{Dept. of Electrical and Computer Engineering}, \orgname{University of California}, \city{Santa Barbara}, \postcode{93106}, \state{CA}, \country{USA}}

\affil[3]{\orgdiv{Quantum Foundry}, \orgname{University of California}, \city{Santa Barbara}, \postcode{93106}, \state{CA}, \country{USA}}

\affil[4]{\orgdiv{Dept. of Chemical Engineering and Materials Science}, \orgname{University of Minnesota}, \city{Minneapolis}, \postcode{55455}, \state{MN}, \country{USA}}

\affil[5]{\orgname{National Institute of Standards and Technology}, \city{Gaithersburg}, \postcode{20899}, \state{MD}, \country{USA}}

\affil[6]{\orgdiv{Advanced Light Source}, \orgname{Lawrence Berkeley National Laboratory}, \city{Berkeley}, \postcode{94702}, \state{CA}, \country{USA}}

\affil[7]{\orgdiv{Stanford Synchrotron Radiation Lightsource}, \orgname{SLAC National Accelerator Laboratory}, \city{Menlo Park}, \postcode{94025}, \state{CA}, \country{USA}}

\affil[8]{\orgname{Oakridge National Laboratory}, \city{Oak Ridge}, \postcode{37830}, \state{TN}, \country{USA}}

\affil[9]{\orgdiv{Materials Research Institute}, \orgname{Pennsylvania State University}, \city{University Park}, \postcode{16802}, \state{PA}, \country{USA}}

\affil[10]{\orgdiv{Dept. of Materials Science and Engineering}, \orgname{University of California}, \city{Santa Barbara}, \postcode{93106}, \state{CA}, \country{USA}}

\affil[11]{\orgdiv{Dept. of Materials Science and Engineering}, \orgname{Pennsylvania State University}, \city{University Park}, \postcode{16802}, \state{PA}, \country{USA}}

\abstract{Altermagnets are a newly identified family of collinear antiferromagnets with a momentum-dependent spin-split band structure of non-relativistic origin, derived from spin-group symmetry-protected crystal structures. Among candidate altermagnets, CrSb is attractive for potential applications because of a large spin-splitting near the Fermi level and a high N\'{e}el transition temperature of around 700 K. We use molecular beam epitaxy to synthesize CrSb (0001) thin films with thicknesses ranging from 10 nm to 100 nm. Structural characterization, using reflection high energy electron diffraction, scanning transmission electron microscopy, and X-ray diffraction, demonstrates the growth of epitaxial films with good crystallinity. Polarized neutron reflectometry shows the absence of any net magnetization, consistent with antiferromagnetic order. {\it In vacuo} angle resolved photoemission spectroscopy (ARPES) measurements probe the band structure in a previously unexplored regime of film thickness, down to 10 nm. These ARPES measurements show a bulk-type, three-dimensional momentum-dependent band splitting of up to 0.7 eV with g-wave symmetry, consistent with that seen in prior studies of bulk single crystals. The distinct altermagnetic band structure required for potential spin-transport applications survives down to the $\sim 10$ nm thin film limit at room temperature.}

\maketitle

\section{Introduction}

Altermagnetism refers to a recently identified class of collinear antiferromagnets wherein the crystal symmetry-compensated magnetic order is accompanied by the breaking of parity-time (PT) symmetry ~\cite{PhysRevX.12.040002,PhysRevX.12.040501,cheong2025kinetomagnetismaltermagnetism}. This leads to a spin-split electronic band structure even in the absence of a net magnetization and with an unusual characteristic: the spin splitting reverses sign in different (collinear and non-collinear) directions of momentum space based on d-, g-, or i-wave symmetry of the altermagnets. Unlike the well-known spin-momentum correlation in topological Dirac materials where strong spin-orbit coupling plays a dominant role ~\cite{10.1088/0034-4885/79/9/094504}, the momentum-dependent non-relativistic spin splitting in altermagnets arises instead from crystal rotation symmetries that create a correspondence between real space and momentum space. Theory predicts a variety of potentially interesting electronic, magnetic, and optical properties in altermagnets~\cite{Smejkal_PhysRevX.12.031042}. Additionally, given their antiferromagnetic order, altermagnets have a natural role to play within the context of antiferromagnetic spintronics where the absence of stray fields, robustness against external magnetic fields, and THz spin dynamics create attractive opportunities in spin-based information technologies~\cite{Baltz_RevModPhys.90.015005}. Altermagnetism is a field still in its nascent stage with the synthesis, characterization, and understanding of various candidate altermagnetic materials being actively pursued. Much of this early work on altermagnets, both theoretical and experimental, has focused on the electronic and magnetic behavior of crystals in the bulk regime ~\cite{https://doi.org/10.1002/advs.202406529,PhysRevLett.133.206401,lu2024observationsurfacefermiarcs,li2024topologicalweylaltermagnetismcrsb,Osumi_PhysRevB.109.115102,Yang2025}, even when explored in thin films~\cite{Reimers2024,Krempasky_Nature,Lee_PhysRevLett.132.036702}. The behavior of candidate bulk-type altermagnets in the limit approaching two dimensions remains relatively unexplored~ \cite{Mazin_2D,cuxart2025emergentmagneticstructures2d}. This raises a relevant question that has not yet been experimentally addressed: how thin can one make an altermagnet and still preserve the characteristic band splitting? At what thickness does the influence of reduced dimensionality, quantum confinement, and heterointerfaces become an important perturbation on the bulk band structure? In this paper, we make some progress in answering these questions through the molecular beam epitaxy (MBE) growth, structural analysis, and angle resolved photoemission spectroscopy (ARPES) of epitaxial thin films of a canonical g-wave altermagnet, CrSb. We find that the predicted altermagnetic band structure is observable at least down to thicknesses of 10 nm, setting a lower bound for preserving the predicted bulk band structure.

Thermodynamically, CrSb is stable at an elemental ratio of 1:1 for temperatures close to $\sim$ 720 K, while at higher temperatures, the phase of CrSb$_2$ is more stable~\cite{Xia2018}. Although CrSb is a ferromagnet in a metastable zinc blende structure~\cite{ZHAO2003507}, the hexagonal NiAs-crystal phase is a well-established A-type antiferromagnet with a high N\'{e}el temperature ($T_N \sim 700$) K in the bulk ~\cite{Snow_PhysRev.85.365,D0DT03277H}. The easy axis of the N\'{e}el vector is parallel to the (0001) axis, that is, the Cr spins are oriented perpendicular to the basal plane ferromagnetically and the planes are aligned antiferromagnetically along the c-axis (Fig. 1a)~ \cite{RevModPhys.25.127,PhysRev.129.2008}. The NiAs-phase of CrSb is one of the first theoretically identified canonical bulk-type g-wave altermagnets with a predicted large spin splitting of around 1.2 eV at the Fermi level ~\cite{Smejkal_PhysRevX.12.031042,Yang2025,PhysRevX.12.040501}  
The synthesis of NiAs-type CrSb films predates the notion of altermagnetism and includes the molecular beam epitaxy (MBE) of CrSb (0001) on GaAs (111) substrates using ferromagnetic MnSb as a buffer~\cite{BURROWS20191783} and CrSb within magnetically doped topological insulator heterostructures~\cite{He2017}. 
Motivated by the altermagnetism scenario, recent attention has turned to the growth of different orientations of CrSb thin films via sputtering ~\cite{Reimers2024,Zhou2025,bommanaboyena2025singlecrystallinecrsb0001filmsgrown} on metallic buffer layers and MBE growth of CrSb in the $(\bar{1}10)$ orientation on GaAs (001) substrates with FeSb as buffer ~\cite{aota2025epitaxialgrowthtransportproperties}. ARPES measurements on CrSb have principally focused on bulk crystals ~\cite{https://doi.org/10.1002/advs.202406529,lu2024observationsurfacefermiarcs,li2024topologicalweylaltermagnetismcrsb,PhysRevLett.133.206401,Yang2025} and bulk-like films (thickness greater than 30 nm) ~\cite{Reimers2024}, showing results that agree with theoretical calculations of the bulk band structure. 

Here, we report the synthesis and characterization of CrSb (0001) thin films with thickness ranging from 10 nm - 100 nm. We use a comprehensive suite of characterization techniques to show that the samples are antiferromagnetic epitaxial CrSb films in the NiAs phase. The crystal structure, crystallinity, epitaxial nature, composition, and morphology of the films are assessed using reflection high energy electron diffraction (RHEED), X-ray diffraction (XRD), atomic-resolution high angle annular dark field (HAADF)-STEM measurements, atomic force microscopy (AFM), and X-ray photoemission spectroscopy (XPS).
To confirm the absence of magnetization in our epitaxial thin films, as expected for spin compensated altermagnets, we use polarized neutron reflectivity (PNR) measurements to probe the depth profile of the magnetization. 
ARPES spectra reveal the three dimensional g-wave altermagnetic band structure of CrSb in these thin films. The spectra are of comparable quality to those obtained on bulk cleaved samples~\cite{Yang2025}. We observe a momentum-dependent altermagnetic band splitting of up to 0.7 eV in CrSb films down to 10 nm thickness, setting a lower thickness bound for the preservation of bulk-like altermagnetic band structure in this material.  

\begin{figure}
    \centering
    \includegraphics[width=0.9\textwidth]{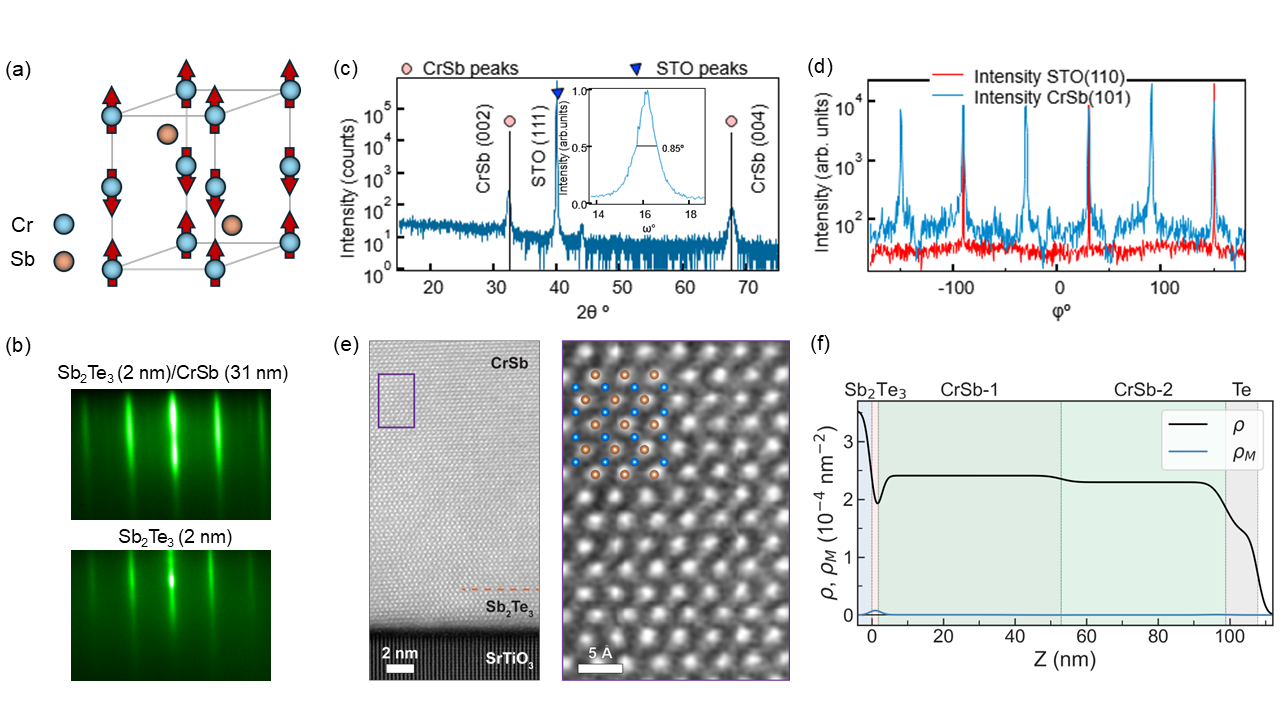}
    \caption{\small \noindent: \textbf{Characterization of an MBE-grown CrSb thin film on SrTiO$_3$ with a thin Sb$_2$Te$_3$ buffer layer}. \textbf{(a)} Schematics of the CrSb unit cell with antiparallel spin sublattices. \textbf{(b)} RHEED patterns of the Sb$_2$Te$_3$ buffer layer and the CrSb film. \textbf{(c)} XRD $2\theta-\omega$ scan of the 31 nm CrSb film with the the rocking curve. \textbf{(d)} XRD phi scan of the SrTiO$_3$ substrate and the CrSb film. \textbf{(e)} HAADF-STEM cross-section image of the Sb$_2$Te$_3$(2 nm)/CrSb(31 nm) heterostructure. Atomic resolution image from the CrSb layer in the purple box (right panel), showing the atomic arrangement of Cr (blue) and Sb (gold). \textbf{(f)} PNR measurements of CrSb(100 nm)/Sb$_2$Te$_3$(2 nm)/SrTiO$_3$ at $T = 300$~K.}
    \label{fig:1}
\end{figure}

\section{Results and discussion}

\subsection{\label{sec:level2}{Epitaxial growth and characterization of CrSb (0001) thin films}}

We grow the CrSb thin films on SrTiO$_3$ (111) substrates in a Scienta Omicron EVO-50 MBE system with a base pressure less than $5 \times10^{-10}$ mBar. To promote the epitaxial growth of CrSb, we first grow a very thin Sb$_2$Te$_3$ ($\sim$2 nm) buffer layer between the substrate and the CrSb film. The anticipated in-plane lattice mismatch between the Sb$_2$Te$_3$ buffer and the CrSb is around 4.3\%. Although CrSb is not a van der Waals (vdW) material, the vdW surface of Sb$_2$Te$_3$ with weak bonding appears beneficial for the epitaxial growth of CrSb. As we show in detail in Supplementary Note 1, the substrate temperature and the Cr-to-Sb beam flux ratios have a significant influence on the CrSb sample quality, including the crystallinity, surface reconstruction, film roughness, and surface topography.  

We first discuss the growth and characterization of a CrSb film with thickness similar to that used in prior reports using sputtering~\cite{Reimers2024}. (A detailed description of the epitaxial growth of CrSb films of 10 nm thickness under varying growth conditions is given in Supplementary Notes 1, 2). The detailed structure we use is SrTiO$_3$/Sb$_2$Te$_3$(2nm)/CrSb(31nm). The growth temperature of this CrSb sample is kept at 240$^{\circ}$C as measured by a thermal camera, with a Cr-to-Sb beam equivalent pressure ratio (BEPR) of 1:5.8 (measured using an ion gauge). The growth of the film is monitored using RHEED. More details of the growth are given in the Methods section. Figure \ref{fig:1}(b) shows the streaky RHEED patterns of the Sb$_2$Te$_3$ buffer layer and the CrSb film, indicating the epitaxial growth of both layers with a relatively smooth surface. Atomic force microscopy (AFM) scans on films grown under optimized conditions show roughness with Rq $\sim 1$~nm over areas of 1 \textmu m$^2$ (see Supplementary Note 1, 2).

$2\theta-\omega$ XRD measurements probe the out-of-plane orientation of the sample and show diffraction peaks consistent with the c-axis of the NiAs phase (Fig.\ref{fig:1}(c)). A small peak due to misoriented CrSb (1100) grains is sometimes seen near 44°. The Sb$_2$Te$_3$ buffer is not resolved in the diffraction pattern. Using the CrSb $(10\bar{1}3)$ reflection, we calculate the lattice constants to be $a = 4.075~\text{\AA}$ and $c = 5.508~\text{\AA}$, indicating $\leq1\%$ compressive in-plane strain, while noting there is some spread in the bulk lattice constants reported in the ICDS PDF5+ database. We also performed XRD $\phi$ scans to investigate the in-plane epitaxial relation of the CrSb with the SrTiO$_3$ substrate, as shown in Fig.\ref{fig:1}(d). The CrSb $(10\bar{1}1)$ $\phi$ scan shows a six-fold symmetry, consistent with the CrSb (0001) film orientation, that is aligned with the 3-fold SrTiO$_3$ (110) planes, confirming a good in-plane epitaxy of the CrSb film with the substrate.  

We further study our thin film quality using cross-sectional high-angle annular dark-field (HAADF) scanning transmission electron microscopy (STEM) imaging. Figure \ref{fig:1}(e)) shows an example from these data. The ${\sim}$ 2 nm region near the interface with the SrTiO$_3$ substrate corresponds to the intended Sb$_2$Te$_3$ buffer layer. We note, however, that many regions of this buffer contain significant diffusion of Cr and a detailed accounting of the resulting crystal structure in this very thin region is challenging (see Supplementary Note 6 for a more detailed discussion). On top of this buffer layer, the HAADF image shows the 31 nm thick CrSb layer with an atomically smooth interface. STEM energy dispersive X-ray (EDX) spectroscopy shows that this CrSb layer has the expected 1:1 concentration ratio of Cr:Sb (see Supplementary Note 6), confirming that we have obtained the desired phase. 

PNR measurements are carried out on thicker (50 nm and 100 nm) CrSb samples at 300 K, well below the $T_N$ for CrSb, in a 1 T magnetic field aligned along the films’ surface. We expect this field to align any uncompensated magnetic moments within the heterostructure into the plane of the film along the field direction. For both CrSb thicknesses, the chosen model that produces the best fit use a uniform nuclear scattering length density (nSLD) characterized by a parameter $\rho$ for each layer, but with the CrSb being split into two sublayers (CrSb-1 and CrSb-2). The fits of the reflectivity data for both samples can be seen in Supplementary Note 7 (Figs. S7 and S8). The reconstructed PNR depth profile of the 100 nm CrSb sample, shown in Fig. 1f, indicates that CrSb-2 has a slightly reduced $\rho$ (2.297 × 10$^{-4}$ nm$^{-2}$) compared to CrSb-1 (2.417 × 10$^{-4}$ nm$^{-2}$), and this difference is supported by the lack of overlap in their 95\% confidence intervals of 2.286 - 2.307 × 10$^{-4}$ nm$^{-2}$ and 2.400 – 2.437 × 10$^{-4}$ nm$^{-2}$, respectively. The respective thicknesses of these two sublayers are 51.3 ± 0.6 nm and 45.6 ± 0.6 nm. For the CrSb 50 nm sample, the difference between the nSLD of CrSb-1 and CrSb-2 is reduced (2.434 × 10$^{-4}$ nm$^{-2}$ and 2.389 × 10$^{-4}$ nm$^{-2}$), with 95\% confidence intervals that nearly overlap. Thus, while using two sublayers does improve the fit, we regard the evidence for multiple sublayers in the 50 nm sample to be marginal. Overall, these findings suggest that a slight reduction in density occurs for thicknesses of CrSb beyond about 50 nm, which is likely caused by reduced substrate-induced strain as layer thickness increases during sample growth.  

Importantly, PNR also probes the magnitude and distribution of magnetization in the structure. This is measured through the magnetic scattering length density (mSLD) characterized by a parameter $\rho_M$. Depth profiles for the chosen best fit model of the 100 nm sample (Fig. 1f) and for the 50 nm sample (Supplementary Note 7, Fig. S8) show the only significant $\rho_M$ contribution (and thus magnetization) is confined to the Sb$_2$Te$_3$ buffer layer. In both the CrSb-1 and CrSb-2 layers, uncertainty analysis of $\rho_M$ shows no significant deviation from zero in the CrSb layer for the 100 nm or the 50 nm CrSb samples (Fig. S8), suggesting no magnetization lies within that layer. Further, models that constrain magnetization to only the CrSb layers and have no $\rho_M$ contribution in the Sb$_2$Te$_3$ layer (Fig. S9), produce poorer quality fits and do not adequately capture the splitting between $R+$ and $R-$ (Supplementary Note 7: Fig. S9 (b), (c)) caused by the magnetization in the samples. The $\rho_M$ in the Sb$_2$Te$_3$ layer is probably caused by Cr diffusion into that thin layer during sample growth, leading to a small net magnetization of 52.9 kA m$^{-1}$ and 32.6 kA m$^{-1}$ for the 100 nm CrSb and 50 nm CrSb samples, respectively. This is consistent with the well-known ferromagnetism induced in (Bi,Sb)$_2$Te$_3$ tetradymite films by Cr doping (see Supplementary Note 6: Fig. S6) \cite{kou2013,ye2015}. However, we note that the Cr interdiffusion-induced ferromagnetism in the samples measured here may be due to some other crystalline phase involving Cr, Sb and Te (see comments in Supplementary Note 6). Regardless of these structural uncertainties regarding the buffer layer, electrical transport measurements clearly show an anomalous Hall effect (AHE) that shows soft hysteresis loops at low temperature and non-linear Hall effect for $T \lesssim 250$~K (see Supplementary Note 5: Fig. S5(b)). This AHE likely arises from the weak magnetization detected in PNR (in a field of 1 T). As the temperature is lowered, it is likely that this magnetization strengthens and contributes to the AHE. Note that the AHE in these CrSb (0001) thin films cannot be attributed to altermagnetism because the three-fold rotation (C$_3$) symmetry around the (0001) axis of the g-wave altermagnet CrSb, precludes this ~\cite{Smejkal_PhysRevX.12.031042,Zhou2025,GUO2023100991}. Consistent with these symmetry-based expectations, we do not observe any AHE at room temperature, where PNR does not show any net magnetization in the CrSb layer.

\subsection{\label{sec:level2}{Band structure of CrSb thin films}}

\begin{figure}
\includegraphics[width=0.9\textwidth]{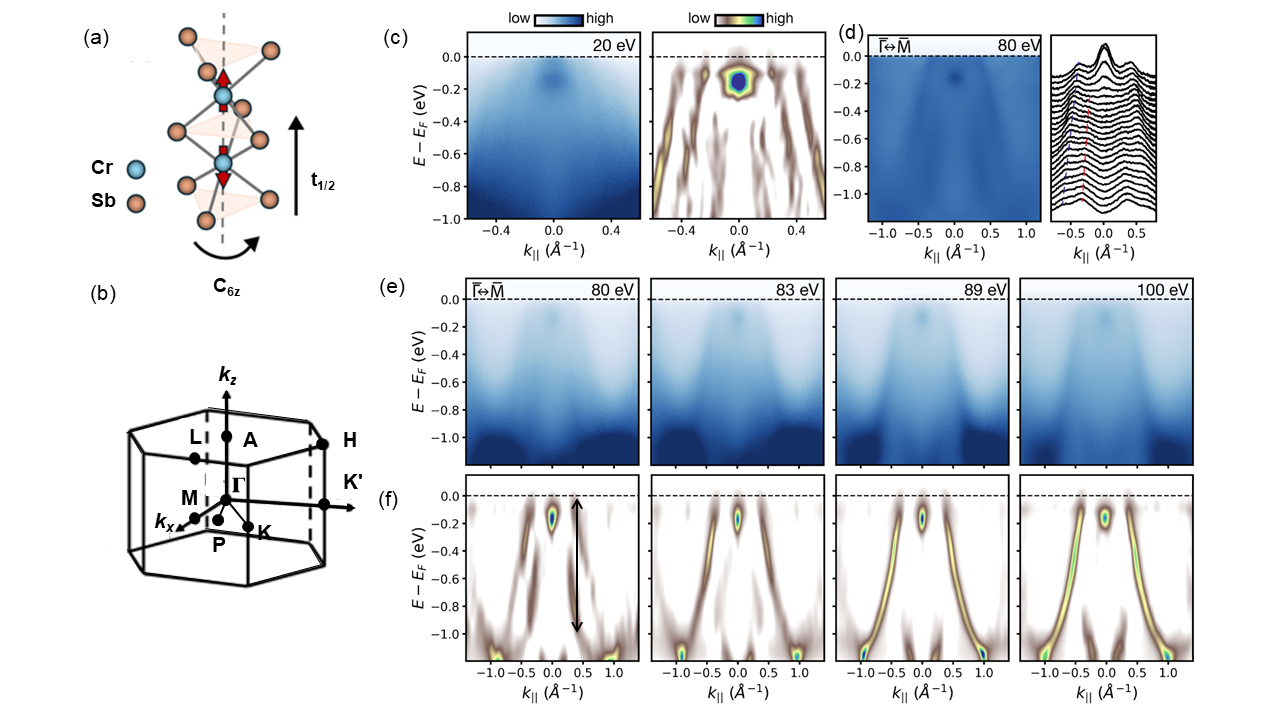}
\caption{\small \noindent\textbf{Photon energy dependent ARPES measurements of 10 nm CrSb thin films.} \textbf{(a)} Schematics of the crystal symmetry and \textbf{(b)} the Brillouin zone of CrSb. \textbf{(c)} $\bar{\Gamma} - \bar{M}$ cut at $h\nu = 20$ eV and corresponding $2D$ curvature plot showing evident band splitting. \textbf{(d)} $\bar{\Gamma} - \bar{M}$ cut at $h\nu = 80$ eV and corresponding MDC showing evident band splitting superimposed on a relatively large background. Blue and red lines are guide to the eye. \textbf{(e)} $\bar{\Gamma} - \bar{M}$ cuts from $h\nu = 80$ eV to $100$ eV. Band splitting disappears closer to $h\nu = 100$ eV which corresponds to the bulk $A$ point. \textbf{(f)} $2D$ curvature plots of the spectra shown in (e) to emphasize splitting. The vertical arrow shows a maximum splitting of 700 meV in the $h\nu = 80$ eV data.
} 
\label{fig:synch}
\end{figure}

ARPES measurements provide insights into the electronic properties of our CrSb films. The CrSb crystal structure consists of two real-space sublattices hosting opposite spins that can be translated by 
non-relativistic spin-group operations such as $[T]$ $[C_{6z}$t$_{1/2}$], where $C_{6z}$ is the six-fold rotational operator around the CrSb (0001) axis, $t_{1/2}$ is the half-unit cell translation operator, and $T$ is the
time-reversal operator, as illustrated in Fig. \ref{fig:synch}(a). First-principles calculations of the spin-dependent band structure in bulk CrSb and experimental ARPES measurements on bulk-cleaved CrSb 
~\cite{PhysRevLett.133.206401,https://doi.org/10.1002/advs.202406529, li2024topologicalweylaltermagnetismcrsb, Yang2025} have shown that CrSb is a bulk-type g-wave altermagnet with significant spin degeneracy of bulk bands on four high symmetry nodal planes in momentum space, such as, $\Gamma- M- K$ and  the three $\Gamma-A-H-K$, and also at the Brillouin zone (BZ) boundary $A-H-L$ (Fig. \ref{fig:synch}b). Away from these high-symmetry planes, substantial spin splitting due to altermagnetism can be identified. To investigate the electronic properties of the g-wave altermagnet in the thin film regime, we probe the three dimensional Brillouin zone of epitaxial CrSb (0001) using synchrotron ARPES and He lamp ARPES. We transfer the samples {\it in vacuo} from the MBE chamber to a connected ARPES chamber and also use a custom built vacuum suitcase operating at $\sim 2 \times 10^{-11}$ Torr to convey samples to the end station.  

We first describe synchrotron-based ARPES measurements on 10 nm thick epitaxial CrSb thin films. These are performed using beamline 10.0.1.2 at the Advanced Light Source and beamline 5-2 at Stanford Synchrotron Radiation Lightsource. The sample temperature during measurements is $\sim 10$ K and the measurements are performed using $p$-polarized light. Tuning the photon energy enables probing momentum space at specific $k_z$ values where the spin-splitting is more obvious along the bulk $\Gamma-A$ direction. Figure \ref{fig:synch} summarizes the synchrotron results. Figure \ref{fig:synch}(c) faintly shows spin splitting in the raw data at $20$ eV, which is away from the high symmetry planes. This spin splitting is more pronounced when looking at the $2D$ curvature. At $80$ eV, Fig. \ref{fig:synch}(d), the spin splitting, with a large value of $\sim 700$ meV is evident in both the raw data and corresponding MDCs, even on top of a large background. Assuming an inner potential, $V_0$ of $17$ eV~\cite{Yang2025}, $80$ eV corresponds to a $k_z$ value halfway between $\Gamma-M-K$ and the BZ boundary $A-H-L$. The exact photon energy corresponding to the bulk high symmetry points likely is different in the thin films since the out of plane lattice constant is different due to the epitaxial nature. In Fig. \ref{fig:synch}(e,f), we show spectra along the $\bar{\Gamma}-\bar{M}$ direction with their corresponding $2D$ curvature plots. As the photon energy is tuned and the corresponding $k_z$ is moved from $\Gamma$ to $A$ the spin splitting is experimentally accessed. Along this momentum space cut, the spin-splitting should be the largest. These results are consistent with previous synchrotron ARPES measurements on bulk CrSb~\cite{PhysRevLett.133.206401,li2024topologicalweylaltermagnetismcrsb,Yang2025,https://doi.org/10.1002/advs.202406529}. Thus, our photon energy-dependent measurements on CrSb thin films reveal the presence of bulk-like split bands even in the thin film regime of 10 nm, consistent 
with theoretical predictions.

\begin{figure}
\includegraphics[width=0.9\textwidth]{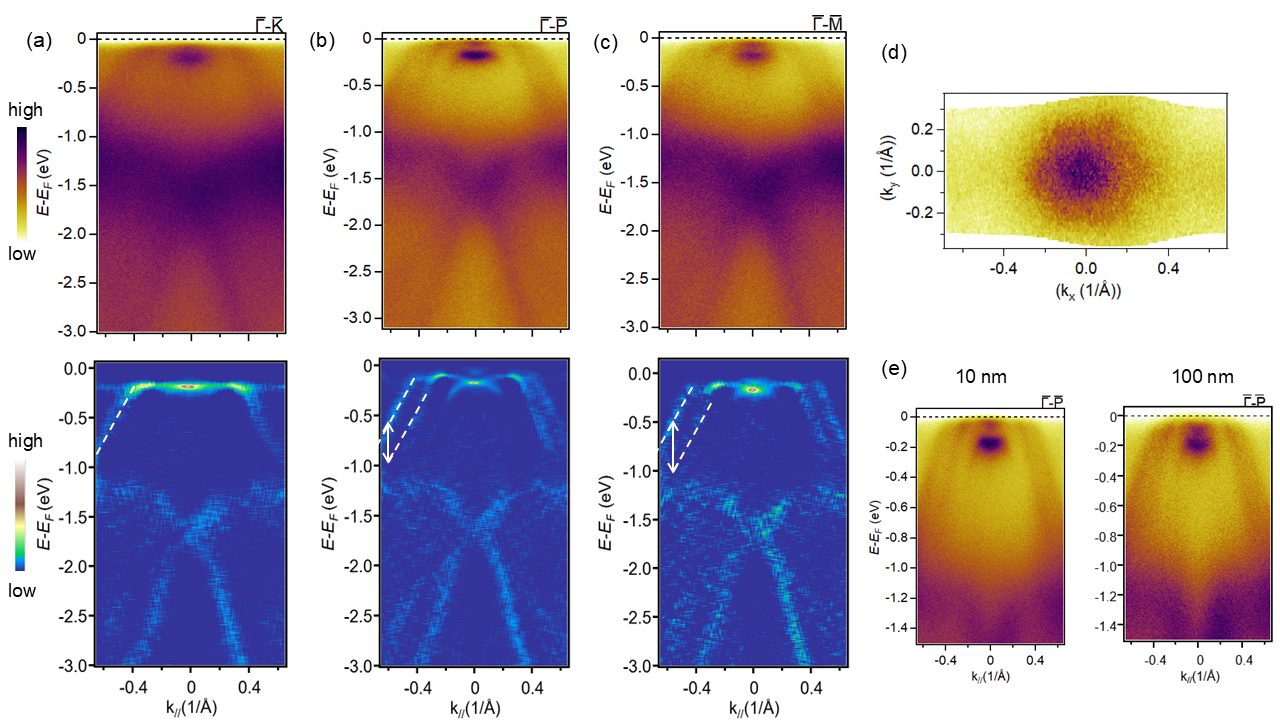}
\caption{\small: \textbf{ He lamp ARPES measurements of CrSb thin films.}  ARPES spectra of CrSb films (10 nm thickness) along different in-plane momenta direction \textbf{(a)}, \textbf{(b)}, \textbf{(c)} (top) and corresponding $2D$ curvature images (bottom). The color lines are guides to the eyes for the split bands in CrSb. Fermi surface map \textbf{(d)} of CrSb (10 nm thickness), \textbf{(e)} comparison of the split bands in CrSb films of 10 nm and 100 nm thickness.} 
\label{fig:arpes}
\end{figure}

The synchroton-based ARPES spectra discussed above confirm the altermagnetic band splitting at low temperature ($T \sim 10$ K). We now discuss ARPES measurements that probe the presence of altermagnetic band splitting in 10 nm epitaxial CrSb samples at room temperature ($\sim$ 300 K). These measurements also aim to identify the in-plane symmetry of the band spectrum. After transferring the thin films {\it in vacuo} from the MBE growth chamber to a UHV connected ARPES chamber, we use the 21.2 eV $I \alpha$ spectral line from a helium plasma lamp as excitation and the photoelectrons are detected by a Scienta Omicron DA 30L analyzer with 6 meV energy resolution. From the synchrotron measurements (Fig. \ref{fig:synch}(c)), we know that the 20 eV photon energy measurements probe a $k_z$ value away from the high symmetry planes. To investigate the in-plane symmetry of the predicted g-wave altermagnet CrSb, we perform systematic measurements of band spectra with the momentum $k_{\parallel}$ axis along the $\bar K-\bar \Gamma- \bar K$, $\bar P- \bar \Gamma- \bar P$, and $\bar M- \bar \Gamma- \bar M$ directions (see Fig. \ref{fig:synch}(b)). We use three different epitaxial 10 nm thin films, grown at optimum growth conditions (see Supplementary note 1, 2), to probe each direction. Near the Fermi level, bands closer to $k_{\parallel}=0$ are visible with splitting signatures dependent on the momentum direction consistent with spin split bands predicted and observed in CrSb bulk crystals~\cite{PhysRevLett.133.206401,li2024topologicalweylaltermagnetismcrsb,Yang2025,https://doi.org/10.1002/advs.202406529}. We observe minimal or vanishing splitting along the $\bar K- \bar \Gamma- \bar K$ direction (fig. \ref{fig:arpes}(a) top), more clearly visible in the $2D$ curvature plot (Fig. \ref{fig:arpes}(a) bottom). When the momentum $k_{\parallel}$ axis is along the $\bar P- \bar \Gamma- \bar P$, a low symmetry direction between $\bar K- \bar \Gamma- \bar K$ and $\bar M- \bar \Gamma- \bar M$ (Fig. \ref{fig:synch}(b)) shows a splitting of ${\sim}$350 meV (Fig. \ref{fig:arpes}(b)) and $\bar M- \bar \Gamma- \bar M$, shows a maximum of ${\sim}$ 500 meV splitting (Fig. \ref{fig:arpes}(c). For energy regions further below, in all three momenta directions, the higher intensity of spectral weights indicates the appearance of multiple bands around $E_B=-1.5$ eV. Another clear bulk band is identified between $E_B=-1.5$ eV and $-3.0$ eV. Figure \ref{fig:arpes}(d) shows the hexagonal Fermi surface map of the NiAs-type CrSb thin film, confirming that, based on the symmetry of the Brillouin zone, there are three high symmetry nodal lines along the $\bar K- \bar \Gamma- \bar K$. Figure \ref{fig:arpes}(e) compares the low spectral intensity bands with splitting signatures along $\bar \Gamma- \bar P$ in a 10 nm and a 100 nm film. The bands are similar in both films and also comparable to the bulk crystal measurements~\cite{li2024topologicalweylaltermagnetismcrsb,Yang2025} with no additional bands, indicating the absence of quantum confinement effects in the 10 nm film. The fermi level is likely to be shifted upward in binding energy, as indicated by the presence of the high spectral intensity spot at $k_{\parallel}=0$ different from the bulk crystals. Prior studies of bulk CrSb~\cite{Yang2025} predict that the band splitting arises mainly from the Cr orbitals, with insignificant contribution from spin-orbit coupling. This is consistent with the non-relativistic nature of the band splitting in CrSb, promising a potentially different route for electronic and spintronic applications compared with more extensively explored approaches that exploit strong spin-orbit coupling.

\section{Conclusions}

In summary, we have developed well-characterized epitaxial thin films of the canonical {\it g}-wave altermagnet CrSb, grown by MBE on SrTiO$_3$ substrates. Comprehensive characterization of these films shows that they are antiferromagnets in the correct NiAs phase. ARPES measurements using photon energies from 20 eV - 100 eV show the characteristic momentum-dependent band splitting predicted for bulk g-wave altermagnetic phase of CrSb is preserved down to 10 nm film thickness, setting a lower bound for incorporating altermagnetic CrSb films in potential spintronic and other device applications. The large (700 meV) band splitting in these thin films augurs well for potential spintronic applications, assuming that additional steps are taken to reduce the g-wave symmetry somewhat so as to allow for observable spin transport~\cite{Zhou2025}. Although we have also explored thinner CrSb films, the clarity of the ARPES data is at present insufficient to draw conclusions about the detailed band structure. This is likely constrained by the crystalline disorder in CrSb films thinner than 10 nm, at least using the buffer/substrate combination explored here. Theoretical calculations of the band structure of CrSb in the sub-10 nm regime would provide valuable impetus for further experiments aimed at improving the crystalline quality of CrSb films in the quasi-2D limit where the interplay between altermagnetism, quantum confinement, reduced dimensionality, and vicinal interfaces creates a rich playground for engineering spin polarized band structures.  

\section*{Experimental Methods}

\noindent{\bf{Sample growth}}

\smallskip

The CrSb/Sb$_2$Te$_3$ thin film heterostructures were deposited using MBE in a Scienta Omicron EVO50 system under ultrahigh vacuum ($<5\times 10^{-10}$ mbar). The commercially obtained STO (111) substrates (MTI, Shinkosha) were cleaned in an ultrasonicator using acetone, isopropanol and deionized water for 2 minutes each. The substrates were then chemically treated in deionized water at 90$^{\circ}$C for 45 minutes and then in a 25\% HCl solution at room temperature for another 45 minutes. After cleaning, the substrates were annealed for 3 hours under a flow of oxygen in a tube furnace at 980$^{\circ}$C. The treated substrates were outgassed in the MBE chamber at 600$^{\circ}$C for 1 h to clean the surface before the deposition of the thin films. The 2 nm buffer Sb$_2$Te$_3$ was grown at 200$^{\circ}$C at a growth rate of 0.2 nm/min, with Te sublimated at a significant overpressure compared to Sb. For most growths, the initial buffer layer is desorbed by heating the substrate to 350$^{\circ}$C and then followed by depositing the 2 nm buffer Sb$_2$Te$_3$ again at 200$^{\circ}$C. This additional step was found to increase the substrate quality and hence the interface.  The epitaxial 30 nm thick CrSb was grown at a substrate temperature of 240$^{\circ}$C via co-evaporation of Cr (purity: 99.997\%., Alfa Aesar) and Sb (purity: 99.999\%., Thermo Scientific) respectively, with a flux ratio of 1:5.80 and a growth rate of 0.2 nm/min. The 10 nm thin CrSb samples were grown at a slower growth rate of 0.04 nm/min. The outgassing and growth temperatures were measured by an infrared camera and RHEED was monitored using a 13 keV electron gun. 

\smallskip
\smallskip
\noindent{\bf{Angle Resolved Photoemission Spectroscopy}}

\smallskip

{\it In vacuo} ARPES measurements at 300 K were carried out using a helium lamp, with a photon energy of 21.2 eV. The photoelectrons emitted were detected by a Scienta Omicron DA30L analyzer.

Synchrotron ARPES measurements were taken at beamline 5-2 at the Stanford Synchrotron Radiation Lightsource (SSRL) and at beamline 10.0.1.2 at the Advanced Light Source (ALS). Data was taken using p-polarized light and a Scienta Omicron DA30L detector. The sample temperature was ${\sim}$10 K during measurement and the base pressure of the ARPES chamber was lower than $3\times10^{-11}$ Torr. The samples were transferred from the growth systems at the Pennsylvania State University to the University of California Santa Barbara, SSRL and ALS for measurement using a custom designed ultra-high vacuum suitcase with base pressure lower than $4\times10^{-11}$ Torr.

\smallskip
\smallskip

\noindent{\bf{Atomic Force Microscopy}}

\smallskip

{\it Ex situ} topography was measured at 300 K using a Bruker Dimension Icon Atomic Force Microscopy instrument. We used a tip scanning AFM operated in the peak force tapping mode.

\smallskip
\smallskip

\noindent{\bf{STEM Characterization}}

\smallskip
\smallskip

An FEI Helios Nanolab G4 dual-beam Focused Ion Beam (FIB) system with 30 keV Ga ions was used for making cross-section samples for the STEM study. Damaged surface layers were removed using ion-milling at 2 keV and amorphous C and Pt were deposited on the surface to protect from damage on exposure to the ion beam.  STEM experiments were carried out on an aberration-corrected FEI Titan G2 60–300 (S)TEM microscope, with a CEOS DCOR probe corrector, monochromator, and a super-X energy dispersive X-ray (EDX) spectrometer. A probe current of 120 pA and operation voltage of 200 keV were used for operating the microscope, and HAADF-STEM images were acquired with the probe convergence angle of 18.2 mrad with inner and outer collection angles of 55 and 200 mrad in the detector respectively. Bruker Esprit software was used to acquire and analyze EDX elemental maps. 

\smallskip
\smallskip

\noindent{\bf{Electrical Transport Characterization}}

\smallskip

We performed electrical transport measurements in a Quantum Design 
DynaCool Physical Properties Measurement System (PPMS). A mechanically defined Hall bar configuration with lateral dimensions of 1 mm length × 0.5 mm wide was used. 

\smallskip
\smallskip

\noindent{\bf{X-Ray diffraction}}

\smallskip

X-ray diffraction patterns were collected on a 320.00 mm radius Malvern Panalytical X’Pert3 MRD four circle X-ray diffractometer equipped with a line source [Cu K-$\alpha$ 1-2 (1.5405980/ 1.5444260Å)] X-ray tube at 45.0 kV and 40.0 mA. The incident beam path included a 2xGe(220) asymmetric hybrid monochromator with a 1/4° divergence slit. A PIXcel3D 1x1 detector operating in receiving slit mode was used with an active length of 0.5 mm. PHD lower and upper levels were set at 4.02 and 11.27 keV respectively. $\phi$ scans were performed on the same instrument using the PIXcel3D 1x1 detector operating in open detector mode. Analysis was carried out using Jade® software (version 9.1) from Materials Data Inc. (MDI) and the International Centre for Diffraction Data (ICDD) PDF5 database.

\smallskip
\smallskip

\noindent{\bf{Polarized Neutron Reflectivity}}

\smallskip

Structural and magnetic properties of Te (10 nm)/CrSb (x nm)/Sb$_2$Te$_3$ (2 nm)/ SrTiO$_3$ (111) films (x = 50 nm and 100 nm) were probed by PNR experiments on the MAGREF reflectometer at the Spallation Neutron Source at Oak Ridge National Laboratory. Measurements of the 1 cm$^2$ films were taken at 300 K in a 1 T field using a spin-polarized neutron beam produced by a two-mirror v-design transmission supermirror polarizer, with a wavelength band $\lambda$= 0.354 – 0.942 nm. Using a spin flipper, the up (+) or down (-) spin states of the incident neutrons were selected,and measurements of the scattered beam were collected at seven sample angles in specular alignment with the detector. Only the spin-up (R+) and spin-down (R-) cross-sections of the reflectivity were collected as a function of the wavevector transfer normal to the film surface, Q, as any magnetic moments in the film plane are expected to be aligned parallel to the 1 T field.  

PNR provided information on the structural and magnetic properties of the system by probing its nuclear ($\rho$) and magnetic ($\rho_M$) scattering length densities. Modeling and fitting the measured R+ and R- allowed depth profiles of $\rho$ and $\rho_M$ to be reconstructed along the direction normal to the firm surface, Z, extending through the heterostructure to the film-substrate interface. Model degeneracy, where different models produced similar $\chi^2$ values, did not allow unique reconstruction of the sample structure, and required limiting potential models used during the fitting process to those that have physical parameters consistent with those known for the system. To further reduce the possibility of overparameterization and model degeneracy, we fit the data using the simplest model of uniform $\rho$ for all layers and uniform $\rho_M$ across the CrSb layer, and then we increased the complexity of the model, stopping when additional parameters did not substantially improve the quality of the fits. A detailed discussion of how the best fit models were chosen and examples of excluded models can be found in the “Polarized Neutron Reflectometry” section of the Supplemental Material. QuickNXS extraction software was used for the reduction of the measured data, and the Refl1d software package was used for fitting and uncertainty analysis~\cite{quicknxs2025,refl1d2024}.   

\section*{Acknowledgements}

This project was primarily supported by the Penn State Two-Dimensional Crystal Consortium-Materials Innovation Platform (2DCC-MIP) under NSF Grant No. DMR-2039351 (SS, YO, AR, NS). Additional support was provided a seed grant from the Penn State MRSEC Center for Nanoscale Science via NSF award DMR 2011839 (SI, NS), including use of the low-temperature transport facilities (DOI: 10.60551/rxfx-9h58). We also acknowledge NSF Grant No. DMR-2309431 (SG, KAM). Parts of this work were carried out in the Characterization Facility, University of Minnesota, which receives partial support from the NSF through the MRSEC (DMR-2011401). We acknowledge support from the University of California, Santa Barbara (UCSB) National Science Foundation (NSF) Quantum Foundry through Q-AMASE-i Program via Award No. DMR-1906325 (PC, WY, CJP). The research reported here made use of the shared facilities of the Materials Research Science and Engineering Center(MRSEC) at UC Santa Barbara: NSFDMR–2308708 (PC, WY, CJP). This research used resources of the Advanced Light Source, which is a DOE Office of Science User Facility under contract no. DEAC02-05CH11231 (AF). Use of the Stanford Synchrotron Radiation Lightsource, SLAC National Accelerator Laboratory, is supported by the U.S. DOE, Office of Science, Office of Basic Energy Sciences under Contract No. DEAC02-76SF00515 (MH, DL). Certain commercial products or company names are identified here to describe our study adequately. Such identification is not intended to imply recommendation or endorsement by the National Institute of Standards and Technology, nor is it intended to imply that the products or names identified are necessarily the best available for the purpose. 

\section*{Supporting Information}
Supporting Information is available from the Wiley Online Library or from the author.

\section*{Data availability}
The data associated with these samples and shown in the main manuscript as well as in the Supporting Information can be publicly accessed through the 2D Crystal Consortium Lifetime Sample Tracking (LiST) data management system, hosted by scholarsphere.psu.edu at the following DOI: (TBD) .



\end{document}